\documentclass[review]{elsarticle}

\usepackage{amssymb}
\usepackage{amsmath}
\usepackage{graphicx}
\usepackage{color}
\usepackage{geometry}
\usepackage{hyperref}
\usepackage{amssymb}
\usepackage{enumerate}
\usepackage{amsthm,amsfonts,amsmath}
\usepackage{bm}
\usepackage{graphicx}
\usepackage{graphics,epsf,psfrag,pifont,dsfont,bbold}
\usepackage{siunitx}
\usepackage{color}
\usepackage{xspace}
\usepackage{accents}

\newcommand{\id}{{\bm 1}}

\def\b0{\bv{0}}

\newcommand{\be}{\begin{equation}}
\newcommand{\ee}{\end{equation}}
\newcommand{\ben}{\begin{equation*}}
\newcommand{\een}{\end{equation*}}
\newcommand{\bv}[1]{\mathbf{#1}}					

\journal{Journal of \LaTeX\ Templates}

\begin{document}

\begin{frontmatter}

\title{Non-locality and viscous drag effects on the shear localisation in soft-glassy materials}






\author[mymainaddress]{A. Scagliarini \corref{mycorrespondingauthor}}
\author[mymainaddressb]{B. Dollet}
\author[mymainaddress]{M. Sbragaglia}

\cortext[mycorrespondingauthor]{Corresponding author}

\address[mymainaddress]{Department of Physics and INFN, University of ``Tor Vergata'', Via della Ricerca Scientifica 1, 00133 Rome, Italy}

\address[mymainaddressb]{Institut de Physique de Rennes, UMR 6251 CNRS/Universit\'e Rennes 1, Campus Beaulieu, B\^atiment 11A, 35042 Rennes Cedex, France}

\begin{abstract}
We study the Couette flow of a quasi-2d soft-glassy material in a Hele-Shaw geometry. The material is chosen to be above the jamming point, where a yield stress $\sigma_Y$ emerges, below which the material deforms elastically and above which it flows like a complex fluid according to a Herschel-Bulkley (HB) rheology. Simultaneously, the effect of the confining plates is modelled as an effective linear friction law, while the walls aside the Hele-Shaw cell are sufficiently close to each other to allow visible {\it cooperativity} effects  in the velocity profiles (Goyon {\it et al.}, {\em Nature} {\bf 454}, 84-87 (2008)). The effects of cooperativity are parametrized with a steady-state diffusion-relaxation equation for the fluidity field $f=\dot{\gamma}/\sigma$, defined as the ratio between shear rate $\dot{\gamma}$ and shear stress $\sigma$. For particular rheological flow-curves (Bingham fluids), the problem is tackled analytically: we explore the two regimes $\sigma \gg \sigma_Y$ and $\sigma \approx \sigma_Y$ and quantify the effect of the extra localisation induced by the wall friction. Other rheo-thinning fluids are explored with the help of numerical simulations based on lattice Boltzmann models, revealing a robustness of the analytical findings. Synergies and comparisons with other existing works in the literature (Barry {\it et al.}, {\em Phil. Mag. Lett.} {\bf 91}, 432-440 (2011)) are also discussed.
\end{abstract}


\end{frontmatter}


\section{Introduction}\label{sec:intro}

The transition from a liquid to an amorphous solid, also known as a {\it jamming transition}, occurs in a wide variety of systems such as emulsions, foams, and granular materials~\cite{Mason96,Katgert13,Kamrin12,Larson}. Above the jamming point, a yield stress $\sigma_Y$ emerges, below which the material deforms elastically and above which it flows like a complex fluid. Upon confinement and increase of the droplets/bubbles/particles concentration, a challenging question concerns the role of  microscopic plastic rearrangements and the emergence of their spatial correlations exhibiting {\it cooperativity} flow behavior at the macroscopic level~\cite{Goyon08,Goyon10,Jop12,Katgert10,Bocquet09}. Such rearrangements affect the overall rheological behaviour of the material, usually described by the Herschel-Bulkley (HB) law of rheometry, relating the stress $\sigma$ to the shear rate $\dot{\gamma}$. Goyon {\it et al.}~\cite{Goyon08} have demonstrated that a modification of the local continuum theory can be successful in accounting for the observed experimental velocity profiles of concentrated emulsions. In particular, they introduced the concept of a spatial cooperativity lengthscale $\xi$ and postulated that the fluidity, defined as $f = \dot{\gamma}/\sigma$, is proportional to the rate of plastic events~\cite{Bocquet09} and follows a diffusion-relaxation equation when it deviates from its bulk value
\be\label{eq:fluidity_vec}
\xi^2 \Delta f(\bm{r})+ f_b(\sigma(\bm{r}))-f(\bm{r}) =0.
\ee
The quantity $f_{b}$ is the bulk fluidity, i.e. the value of the fluidity in the absence of spatial cooperativity ($\xi=0$). The non-local equation (\ref{eq:fluidity_vec}) has been justified~\cite{Bocquet09} based on a kinetic model for the elastoplastic dynamics of a jammed material, which takes the form of a non-local kinetic equation for the stress distribution function. In the steady state, under the hypothesis of weak cooperativity, the model predicts non-local equations of the form (\ref{eq:fluidity_vec}), plus an equation predicting a proportionality between the fluidity and the rate of plastic events. This picture was later applied to other complex fluids, such as Carbopol gels~\cite{Geraud13}, granular media~\cite{Kamrin12,Amon12}, and foams in a 2d cylindrical Couette geometry~\cite{Katgert10}. The spatial cooperativity was shown to be of the order of a few times (typically five) the size of the elementary microstructural constituents, i.e. the droplets for emulsions~\cite{Goyon08,Goyon10,SOFT14}, the bubbles for foams \cite{Katgert10}, the blobs for a polymeric gel~\cite{Geraud13}.  The fluidity model agrees with existing experiments, and provides a convenient framework to rationalize the flow of confined complex fluids.  However, at least two points remain unclear and largely unexplored. First, the issue of the boundary condition at solid walls for $f$. Only recently, Mansard {\it et al.}~\cite{Mansard14} explored the role of surface boundary conditions for the flow of a dense emulsion. Both slippage and wall fluidization were shown to depend non-monotonously on the roughness. Second, the fluidity parameter $f$ has been seldom related to an independent and direct measure of the local density of plastic events. Sometimes, indirect indications of such a relation have been proposed, based on the correlations of the fluctuations of the shear rate~\cite{Jop12}. Using numerical simulations based on the bubble model~\cite{Durian97}, Mansard {\it et al.}~\cite{Mansard13} were able to measure independently the fluidity and the density of plastic events, but they show that the two quantities are not proportional; more precisely, the rearrangement rate was found to be a sublinear power (with an exponent 0.4) of the fluidity. On the other hand, using experiments in a Hele-Shaw cell and simulations based on lattice Boltzmann method, we showed recently \cite{ourJFM} that for foams and emulsions flowing in a 2d channel, there is a good correlation between the rate of plastic events and the fluidity. \\
Very frequently some of the systems of interest are confined so as to be quasi-2d: this is the case of Hele-Shaw cells~\cite{Debregeas01,Desmond13a,Desmond13b}, or quasi-2d systems made of bubbles confined between a plate and a liquid surface~\cite{Wang06}. A friction force due to the presence of one or more confining plates may provide {\it shear localisation} for the velocity profiles on lengthscales which can be of the order of a few bubble/droplet sizes, thereby interfering with the cooperativity lengthscale described above. This extra localisation is usually parametrized with another lengthscale related to the viscosity and wall friction~\cite{Janiaud06}. This naturally poses the question on how to rationalize the coupled role of friction and non-locality. Barry {\it et al.}~\cite{Barry11} combined the non-local constitutive equation for the fluidity field \eqref{eq:fluidity_vec} with the continuum theory of 2d shear localisation for a foam in a Couette Flow~\cite{Janiaud06}. They showed that the localisation length due to friction is increased by cooperativity, and explored the limiting cases of zero and infinite cooperativity length. Due to the generality of their formulation, their analysis may be directly applicable to other complex fluids.\\
The aim of this paper is to complement the results by Barry {\it et al.}~\cite{Barry11} exploring the complex flow of a soft-glassy material in a Hele-Shaw geometry with both friction and non-locality. The problem is tackled analytically for the case of a Bingham fluid, where  we study the two regimes $\sigma \gg \sigma_Y$ and $\sigma \approx \sigma_Y$.  A distinctive feature of our analysis, is to explore those situations where the wall acts as a source of fluidity propagating into the bulk of the system~\cite{Mansard14,Mansard13} and to provide analytical results which remain finite in the limit of zero wall friction (see section \ref{sec:model}). In the second part of the paper, we explore the validity and robustness of the analytical findings by performing numerical simulations of the flow of concentrated 2d emulsions under the effect of a linear friction.\\
The paper is organized as follows: in Sec.~\ref{sec:model} we recall the essential features of the theoretical framework for the problem at hand; in Sec.~\ref{sec:analytics} we derive analytical results for a Bingham fluid; in Sec.~\ref{sec:numerics} we recall the essential features of the numerical model used to perform the numerical simulations, while in Sec.~\ref{sec:results} we compare the numerical results with the analytical predictions of Sec.~\ref{sec:analytics}. Conclusions and implications for further studies are finally discussed in Sec.~\ref{sec:conclusions}.

\section{Problem Statement}\label{sec:model}

In this section we briefly recall the essential features of the fluid-dynamical model we consider for our study. The model considers a steady unidimensional flow in a Hele-Shaw cell with a width $H$ and vanishing inertia. We also neglect end effects and assume that the flow is streamwise invariant. Hence, the flow profile writes: $\mathbf{v} = v(z)\hat{x}$, with $\hat{x}$ the streamwise direction and $\hat{z}$ the spanwise one (with $z \in [-H/2; +H/2]$). We set the velocity at the boundaries such that $v(\pm H/2)= \pm v_w$. Crucial in our model is the balancing between shear forces and friction forces
\be\label{eq:momentumbalancesimple}
\frac{d \sigma(z)}{dz} -\beta v(z)=0 ,
\ee
with $\beta$ a wall friction parameter and $\sigma(z)$ the total shear stress.
Furthermore, the notion of {\it cooperativity} and non-local effects induced by local plastic rearrangements is key for our purposes. The underlying idea is that correlations among plastic events exhibit a complex spatio-temporal scenario: they are correlated at the microscopic level with a corresponding cooperativity flow behavior at the macroscopic level. Plastic events trigger avalanches of such processes in their vicinity and the consequent non-local effects are captured in terms of the effective inverse viscosity, or fluidity, $f(z)=\dot{\gamma}(z)/\sigma(z)$, relating stress to strain rate $\dot{\gamma}(z)$, locally. At the mathematical level, this is translated in the following equation
\be\label{eq:fluidity}
\xi^2 \frac{d^2 f(z)}{d z^2}+[f_{b}(\sigma)-f(z)]=0
\ee
where the scale $\xi$ quantifies the non-locality of the cooperativity within the flow. The quantity $f_{b}$ is the bulk fluidity, i.e. the value of the fluidity in absence of spatial heterogeneities. The bulk fluidity $f_b$ only depends upon the shear stress {\it via} the rheological flow curve. As stressed in the original papers~\cite{Goyon08,Bocquet09,Mansard13}, in fact, the bulk fluidity must be interpreted as the fluidity in absence of non-local effect, as it would be for an HB flow-curve $\sigma = \sigma_Y+ K \dot{\gamma}^a$
($a$ and $K$ are characteristic parameters; in particular, for a Bingham fluid, $a=1$ and $K$ is essentially the
plastic viscosity of the material) homogeneously valid, and it is expressed {\it in terms of the shear stress} as
\be \label{eq:fb}
f_b(\sigma) = \frac{1}{\sigma}\left(\frac{\sigma - \sigma_Y}{K}\right)^{1/a}.
\ee
The bulk fluidity (\ref{eq:fb}) is a constant in absence of wall friction since $\sigma=\mbox{const}$ from equation \eqref{eq:momentumbalancesimple}. Calculating $f_b$ from the velocity profile is obviously wrong, the latter being affected by non-local effects: while the bulk fluidity only depends upon the stress, $f(z)$ depends upon the position in space as predicted by equation \eqref{eq:fluidity}.  Moreover, the solution of the fluidity equation requires boundary conditions, i.e. one has to prescribe the value the fluidity close to the boundaries. Equations \eqref{eq:momentumbalancesimple}-\eqref{eq:fb} are coupled together and analytical solutions cannot easily be found: one has to work out the details in some appropriate asymptotic limits or solve the problem numerically~\cite{Barry11}. However, in the case of a Couette flow with zero friction ($\beta=0$), an exact analytical solution can readily be found. In particular, at fixed shear stress $\sigma$ and with boundary conditions $f(\pm H/2)=f_w$, the expression of the shear rate $\dot{\gamma}(z)$ reduces to~\cite{Goyon10}:
\be\label{eq:fluidityCF}
\dot{\gamma}(z)=\sigma \left\{ f_{b}(\sigma)+[f_w-f_{b}(\sigma)]\frac{\cosh( z/\xi)}{\cosh(H/2\xi)} \right\}
\ee
independently of the HB parameters of the flow-curve in \eqref{eq:fb}. Switching on the friction parameter, already at the level of the Couette flow, makes the problem more challenging: the shear stress is no longer constant and the solutions of Eqs.~\eqref{eq:momentumbalancesimple}-\eqref{eq:fb} depends on the parameters of the flow-curve \eqref{eq:fb}. Our strategy will be to work with a Bingham fluid (i.e. $a=1$ in equation \eqref{eq:fb}), which allows to write exact solvable equations in the two limits $\sigma \gg \sigma_Y$ and $0 < \sigma - \sigma_Y \ll \sigma_Y$. In particular, in those limits, the effect of friction will be quantified exactly on the velocity profiles in such a way that in the limit $\beta \rightarrow 0$ we will recover the solution \eqref{eq:fluidityCF}. Through comparisons with numerical simulations~\cite{ourJFM}, for which the flow-curve fulfills the HB equation with an exponent $a < 1$, we will try to capture what we believe are the ``universal'' features that we are able to prove analytically in the case $a=1$.\\
The study that we propose bears analogies with the work of Barry {\it et al.}~\cite{Barry11}, who combined the local model of Janiaud {\it et al.}~\cite{Janiaud06} with a non-local constitutive equation for the fluidity field in a Couette flow. The authors explored analytically the limit of weak ($\xi \ll H$) and strong ($\xi \rightarrow \infty$) cooperativity.  For small $\xi$ (and close to yield) they predicted the emergence of a localisation length $L_v$ of the velocity profile, an increasing function of both the cooperativity length $\xi$ and the friction length:
\be\label{frictionlength}
L_{\beta}=\sqrt{\frac{K}{\beta}} ,
\ee
through the relation $L_v = \sqrt{\xi^2 + L_{\beta}^2}$, which can be approximated (being $\xi$ small) by $L_v \approx L_{\beta} \left( 1 + \frac{\xi^2}{2 L_{\beta}^2} \right)$, while in the limit $\xi \rightarrow \infty$ an exponential profile is recovered with $L_v$ growing with $L_{\beta}$. To work out these results they also dealt with a Bingham fluid, as we do here. However, there are significant differences with our approach which must be underlined, the first of which is of a conceptual character. In~\cite{Barry11} the cooperativity effects are seen as corrections to the underlying continuum model and the {\it bare} fluidity model results~\cite{Goyon08,Bocquet09} are not recovered in the limit of vanishing viscous drag ($L_{\beta} \rightarrow \infty$); instead, we put ourselves in the --somehow-- complementary perspective of tuning the wall friction ($L_{\beta}$) at a given cooperativity ($\xi$), motivated by the aim of comparing with mesoscopic numerical simulations where the latter is fixed by the fluid physical properties (and it cannot be easily related to the parameters of the numerical model). Furthermore, we will explicitly address both the limit of low (close to yield) and high (far from yield) shear stress, showing that wall friction and non-locality conspire to give the global shear localisation in opposite ways in the two regimes.
Finally, the boundary conditions for the fluidity are different: based on the idea that the fluidity equation looks like a steady-state diffusion equation, Barry {\it et al.} assume an {\it adiabatic} boundary condition at the walls, $f^{\prime}(z=\pm H/2)=0$. This contrasts with our choice of using a Dirichlet-type boundary condition, $f(z=\pm H/2)=f_w$; it is indeed our interest to explore those situations where the wall acts as a source of fluidity propagating into the bulk of the system~\cite{Mansard14,Mansard13}, in the spirit of the works by Bocquet {\it et al.}~\cite{Bocquet09} and Goyon {\it et al.}~\cite{Goyon08}. Generally speaking, Mansard \emph{et al.} \cite{Mansard14} recently proposed a mixed boundary condition: $\mp\xi_{\mathrm{wall}} f'(z = \pm H/2) = f(z = \pm H/2) - f_s(\sigma)$, where $\xi_{\mathrm{wall}}$ is a surface cooperativity length, and $f_s(\sigma)$ is the value of the fluidity at the wall when the fluidity gradient vanishes at the wall. The circumstances at which this boundary condition reduces to an adiabatic-like or a Dirichlet condition remain an open issue.


\section{Asymptotic results for a Bingham fluid} \label{sec:analytics}

In this section we report the analytical solutions for the coupled Eqs.~\eqref{eq:momentumbalancesimple}-\eqref{eq:fb} upon the assumption of a Bingham fluid ($a=1$) and explore separately the two regimes, $\sigma \gg \sigma_Y$ ({\it fluid regime}) and $0 < \sigma - \sigma_Y \ll \sigma_Y$ ({\it plastic regime}).

\subsection{Fluid regime ($\sigma \gg \sigma_Y$)}\label{subsec:1}

In this case the bulk fluidity $f_b$ is a constant and equal to the inverse plastic viscosity $K^{-1}$. Under this assumption, the fluidity equation \eqref{eq:fluidity} decouples from that of the velocity profile and can be integrated directly to give
\be \label{eq:fsol}
f(z) =f_b + \frac{f_w -  f_b}{\cosh(\lambda)}\cosh \left(\frac{z}{\xi}\right) ,
\ee
where $\lambda \equiv H/2\xi$. The force balance~\eqref{eq:momentumbalancesimple} can be recast, upon derivation with respect to the variable $z$ and recalling the fluidity $f(z)=\dot{\gamma}(z)/\sigma(z)$, in the following form
\be\label{eq:sigmaloc}
\sigma^{\prime\prime}- \beta f \sigma = 0.
\ee
Inserting the expression (\ref{eq:fsol}) for $f$ we get
\be\label{eq:step0}
\sigma^{\prime\prime} - \beta \left(  f_b + \frac{f_w -  f_b}{\cosh(\lambda)}\cosh \left(\frac{z}{\xi}\right) \right) \sigma = 0.
\ee
Equation \eqref{eq:step0} can be rewritten (upon the change of variable $z \rightarrow \tilde{z} = z/\xi$) as
\be \label{eq:mathieu}
\sigma^{\prime\prime} - \epsilon^2 \left(1-\frac{1 -K f_w}{\cosh(\lambda)}\cosh(\tilde{z}) \right) \sigma = 0,
\ee
where we have defined $\epsilon = \xi / L_{\beta}$ and used the definition of the friction length given in \eqref{frictionlength}. A solution of equation \eqref{eq:mathieu} is
$$
\sigma(\tilde{z}) = \mathcal{M} (\epsilon^2(\beta), q(\beta); 2i \tilde{z})
$$
where $\mathcal{M}$ is the modified Mathieu's function \cite{AbramowitzStegun} and $q=\frac{\epsilon^2}{2}\frac{1-Kf_w}{\cosh(\lambda)}$. If $H\gg\xi$, interesting insight close to the wall $z = H/2$ is provided by the asymptotic limit $\tilde{z} \gg 1$ in equation \eqref{eq:mathieu}. Then we approximate $\cosh\tilde{z} \approx e^{\tilde{z}}/2$, and equation \eqref{eq:mathieu} reduces to
$$
\sigma^{\prime\prime}- \epsilon^2 \left(\frac{K f_w - 1}{2\cosh(\lambda)}e^{\tilde{z}}  + 1 \right) \sigma= 0,
$$
or also, with the change of variable $e^{\tilde{z}/2} = \eta$, to
$$
\eta^2 \sigma^{\prime\prime} + \eta \sigma^{\prime} - 4\epsilon^2 \left(\frac{K f_w - 1}{2\cosh(\lambda)}\eta^2  + 1 \right) \sigma = 0.
$$
Finally, setting $x=\epsilon\eta \sqrt{2(Kf_w - 1)/\cosh(\lambda)}$ and $\alpha = 2\epsilon$, we get
\be \label{eq:bessel}
\sigma^{\prime\prime} + \frac{1}{x} \sigma^{\prime} - \left(1 + \frac{\alpha^2}{x^2} \right)\sigma=0
\ee
which is the Bessel's modified equation. The velocity is positive, hence from (\ref{eq:momentumbalancesimple}), $\sigma$ is a monotonously growing function. A solution of (\ref{eq:bessel}) is then proportional to the modified Bessel function of the first kind $I_{\alpha}(x)$, i.e.
$\sigma(z) = \sigma_w^{(0)}I_{2\xi/L_{\beta}}\left(\epsilon \sqrt{2(Kf_w - 1)/\cosh(\lambda)} \, e^{z/2\xi}\right)$,
where $\sigma_w^{(0)}$ is the stress in absence of wall friction.
The solution for the velocity profile (close to the top wall $z=+H/2$) is, therefore,
\be \label{vprof}
v(z) = v_w + \sigma_w^{(0)} \int_{H/2}^z \left(f_b + \frac{f_w- f_b}{\cosh(\lambda)}\cosh \left(\frac{\zeta}{\xi}\right) \right) I_{2 \xi/L_{\beta}} \left(\frac{\xi}{L_{\beta}}\sqrt{\frac{2(Kf_w -1)}{\cosh(\lambda)}}e^{\zeta/2\xi} \right)d\zeta.
\ee
For small $\xi/L_\beta$ the argument of the Bessel's function is small and we can expand it as $I_{\nu}(x) \sim (x/2)^{\nu}/\Gamma(\nu + 1)$ \cite{AbramowitzStegun}, which gives
\be
v(z)  \sim v_w + \sigma_w^{(0)} \frac{\left(\frac{\xi}{L_{\beta}}\right)^{2\xi/L_{\beta}}\left(Kf_w - 1\right)^{\xi/L_{\beta}}}{K\Gamma\left(\frac{2\xi}{L_{\beta}}+1\right)}\int_{H/2}^z \left[1 + (Kf_w - 1)e^{(\zeta-H/2)/\xi}\right] [e^{(\zeta-H/2)/2\xi}]^{2\xi/L_{\beta}}d\zeta,
\ee
where $\Gamma(x)$ is Euler's Gamma function and we have made use of $f_b \approx K^{-1}$.
The latter equation can be easily integrated resulting in the following expression for the velocity profile
\be \label{eq:velocity}
v(z)\sim v_w + \mathcal{A}\left\{ L_{\beta} \left[e^{(z-H/2)/L_{\beta}} - 1 \right]+ \left(Kf_w - 1 \right) L_v \left[e^{(z-H/2)/L_v} - 1\right]\right\},
\ee
where the coefficient $\mathcal{A}$ is given by
\be \label{eq:Acoeff}
\mathcal{A} = \sigma_w^{(0)}\frac{\left(\frac{\xi}{L_{\beta}}\right)^{2\xi/L_{\beta}}\left(Kf_w - 1\right)^{\xi/L_{\beta}}}{K\Gamma\left(\frac{2\xi}{L_{\beta}}+1\right)}
\ee
and the localisation length $L_v$ by
\be \label{eq:loclength}
L_v = \frac{L_{\beta}\xi}{L_{\beta} + \xi}=\frac{\xi}{1+\xi/L_{\beta}}.
\ee
Equation (\ref{eq:velocity}) suggests that the velocity profile is the result of the superposition of two exponentials with characteristic lengths $L_{\beta}$ and $L_v$. For not too high friction (small $\beta$), $L_{\beta}$ is large and the velocity localisation is controlled by the second exponential, i.e. it is determined by the localisation length $L_v$:
\be \label{eq:velocity2}
v(z) \sim v_w + \mathcal{A} (Kf_w - 1) L_v \left[e^{(z-H/2)/L_v} - 1\right].
\ee
We notice from equation \eqref{eq:loclength} that $L_v$ tends to $\xi$ when $L_{\beta} \rightarrow \infty$ (that is $\beta \rightarrow 0$), as one would expect. Also, for a finite $\xi$, $L_v$ is always smaller than $\xi$: wall friction, then, adds up as an extra source of localisation for the velocity.\\
It is worth commenting that the above scenario could also be predicted based on heuristic arguments. Indeed, equation \eqref{eq:sigmaloc} suggests a stress localisation scale related to $L_{\beta}$, although the exact analytical solution hinges on the knowledge of the function $f=f(z)$. Once a localisation for the stress has been predicted,  it is then straightforward to derive the resulting localisation for the velocity $v(z) = \int \sigma(\zeta) f(\zeta) d\zeta$.

\subsection{Plastic regime ($0 < \sigma - \sigma_Y \ll \sigma_Y$)}\label{subsec:2}


This is also the regime considered by Barry {\it et al.} in~\cite{Barry11} (see equation (12) of their paper) to derive the result in the weak cooperativity limit. In this plastic regime, the effect of friction must be expected to be small: at yield, in fact, the bulk fluidity goes to zero as well as the velocity (the friction force goes as $\sim \beta v$). If we write $\sigma = \sigma_Y + \tilde{\sigma}$, with $\tilde{\sigma} \ll \sigma_Y$, the bulk fluidity (to first order in $\tilde{\sigma}/\sigma_Y$) reads:
\be \label{eq:bfluid}
f_b \simeq \frac{\tilde{\sigma}}{K\sigma_Y}.
\ee
If we now derive the fluidity equation (\ref{eq:fluidity}) twice with respect to $z$, we get
\be \label{eq:fluid4}
\frac{d^4 f}{d z^4} = \frac{1}{\xi^2}\left(\frac{d^2 f}{d z^2}-\frac{d^2 f_b}{d z^2} \right).
\ee
For the second derivative of $f_b$ we see from equation \eqref{eq:bfluid} and from the mechanical equilibrium condition $\sigma^{\prime} = \beta v$ that the following relations hold
$$
\frac{d^2 f_b}{d z^2}= \frac{1}{K\sigma_Y}\frac{d^2 \tilde{\sigma}}{d z^2}= \frac{1}{K \sigma_Y} \frac{d}{d z}(\beta v) = \frac{1}{K\sigma_Y}(\beta \dot{\gamma}),
$$
but $\dot{\gamma} = \sigma f \simeq \sigma_Y f$ (again, to first order
in $\tilde{\sigma}$), hence
$$
\frac{d^2 f_b}{d z^2} \approx \frac{\beta}{K} f
$$
and equation \eqref{eq:fluid4} becomes
\be
\frac{d^4 f}{d z^4} = \frac{1}{\xi^2}\left(\frac{d^2 f}{d  z^2}-\frac{\beta}{K}f \right),
\ee
which is a closed linear fourth-order differential equation for $f$. Using $L_{\beta} = \sqrt{K/\beta}$, the latter equation can be rewritten as
$$
f^{IV} - \frac{1}{\xi^2}f^{\prime \prime} + \frac{1}{\xi^2  L_{\beta}^2}f =0,
$$
whose solution, due to symmetry reasons (i.e. $f(-z) = f(z)$), is
\be \label{eq:fluidb1}
f(z) = C_1 \cosh \left(\frac{z}{L_+}\right) + C_2 \cosh \left(\frac{z}{L_-}\right) ,
\ee
where $C_{1,2}$ are two integration constants and $L_{\pm}$ are such that
\be \label{eq:fluidb2}
\frac{1}{L_{\pm}^2} = \frac{1}{2 \xi^2}\left(1 \pm \sqrt{1 - \frac{4 \xi^2}{L_{\beta}^2}} \right);
\ee
the latter equation provides, in the low friction limit ($\xi \ll L_{\beta}$, to first order in $\xi^2/L_{\beta}^2$),
$L_- \simeq L_{\beta}$ and $L_+ \simeq L_v$, where
\be \label{eq:L0}
L_v = \xi \left( 1 + \frac{\xi^2}{2 L_{\beta}^2} \right),
\ee
whence, equation (\ref{eq:fluidb1}) can be rewritten in the following form, fulfilling the boundary condition
$f(\pm H/2) = f_w$,
\be\label{eq:fluidbeta}
f(z) = f^{(eff)}_b(z; \beta) + \frac{f_w - f^{(eff)}_b((H/2); \beta)}{\cosh  (H/2L_v)}\cosh \left(\frac{z}{L_v}\right) ,
\ee
where (superscript $(0)$ refers to the $\beta \rightarrow 0$ limit)
\be \label{eq:bfluidbeta}
f^{(eff)}_b(z; \beta) = f^{(0)}_b \cosh \left( \frac{z}{L_{\beta}} \right).
\ee
As in the previous section, the zero-friction ($L_{\beta} \rightarrow \infty$) limit of equation \eqref{eq:L0} gives  $L_v \rightarrow \xi$, i.e. we recover the purely cooperative case. Equation (\ref{eq:fluidbeta}) is quite elegant since it has the form of the fluidity without wall friction and it unveils that the effect of the latter is to renormalize the bulk fluidity into an effective one according to equation \eqref{eq:bfluidbeta}. For very large $L_{\beta}$ (i.e. small friction) we can assume, in equation \eqref{eq:fluidbeta}, $f^{(eff)}_b(z; \beta) \approx f^{(0)}_b$, that is the fluidity (and hence the velocity) profile is controlled solely by $L_v$. Equation \eqref{eq:L0} elucidates well the interplay of cooperativity and friction, showing that the localisation length $L_v$ is indeed proportional to the cooperativity length $\xi$ and supports a relative increase proportional to $\xi^2/L_{\beta}^2$. These results are in qualitative agreement with those found by Barry {\it et al.}~\cite{Barry11}, in that wall friction and non-locality conspire to give the global shear localisation. However, in Barry {\it et al.}~\cite{Barry11} the cooperativity is taken as a perturbation to the underlying wall friction and the result tends to diverge at vanishing friction, whereas the localisation length must remain finite in this limit and equal to $\xi$.
As mentioned above these effects are small, namely of second order in $\xi/L_{\beta}$: for comparison, let us recall that in the fluid regime the correction to the localisation length with respected to the no-friction reference case was of the first order. The numerics (see Sec.~\ref{sec:results}) will actually confirm these observations. \\
To conclude this section, let us remark an important result: the velocity profiles exhibit different forms
(controlled by different localisation lengths) in the two regimes, hence they cannot overlap upon simply rescaling
by the wall velocity (i.e. they are shear-rate dependent). Of course, this is an effect due to the conspiring role
of wall friction and cooperativity; in absence of spatial heterogeneities (which translates, in the language
of the kinetic elasto-plastic model \cite{Bocquet09}, into the condition $\xi \rightarrow 0$) the only relevant
length would be $L_{\beta}$ and it is easy to realize that the profiles would recover the rate-independency, as
observed experimentally by comparison of monodisperse and polydisperse 2d foams under linear shear
\cite{Katgert08}. Analogously, without wall friction (i.e. in the limit $L_{\beta} \rightarrow \infty$), $L_v$ tends
to $\xi$, as previously commented, and, again, rate-independent profiles are recovered \cite{SOFT14}.

\section{Numerical Model} \label{sec:numerics}

For the numerical simulations,  we adopt a mesoscopic lattice Boltzmann (LB) model for non ideal binary fluids, which combines a small positive surface tension, promoting highly complex interfaces, with a positive disjoining pressure, inhibiting interface coalescence. The model has already been described in several previous works~\cite{ourJFM,CHEM09,BenziEPL}. Here, we just recall its basic features. We consider two fluids $A$ and $B$, each described by a {\it discrete} kinetic distribution function $f_{\zeta i}({\bm r},{\bm c}_i;t)$, measuring the probability of finding a particle of fluid $\zeta =A,B$ at position ${\bm r}$ and discrete time $t$, with discrete velocity ${\bm c}_i$, where the index $i$ runs over the nearest and next-to-nearest neighbors of ${\bm r}$ in a regular 2d lattice~\cite{CHEM09,BCS}. The distribution functions evolve in time under the effect of free-streaming and local two-body collisions, described by a relaxation towards a local equilibrium ($f_{\zeta i}^{(eq)}$) with a characteristic time scale $\tau_{LB}$:
\be
\label{LB}
f_{\zeta i}({\bm r}+{\bm c}_i,{\bm c}_i;t+1) -f_{\zeta i}({\bm r},{\bm c}_i;t)  = -\frac{1}{\tau_{LB}} \left(f_{\zeta i}-f_{\zeta i}^{(eq)} \right)({\bm r},{\bm c}_i;t)+F_{\zeta i}({\bm r},{\bm c}_i;t).
\ee
The equilibrium distribution is given by
\be
f_{\zeta  i}^{(eq)}=w_i \rho_{\zeta} \left[1+\frac{{\bm v} \cdot {\bm c}_i}{c_s^2}+\frac{{\bm v}{\bm v}:({\bm c}_i{\bm c}_i-c_s^2 \id)}{2 c_s^4} \right] ,
\ee
with $w_i$ a set of weights known a priori through the choice of the discrete velocity set~\cite{Shan06} and $c_s^2=1/3$ a characteristic velocity (a constant in the model). Coarse-grained hydrodynamical densities are defined for both species $\rho_{\zeta }=\sum_i f_{\zeta i}$ as well as a global momentum for the whole binary mixture ${\bm j}=\rho {\bm v}=\sum_{\zeta , i} f_{\zeta i} {\bm c}_i$, with $\rho=\sum_{\zeta} \rho_{\zeta}$. The term $F_{\zeta i}({\bm r},{\bm c}_i;t)$ is just the $i$-th projection of  the total internal force which includes a variety of interparticle forces. First, a repulsive ($r$) force with strength parameter ${\cal G}_{AB}$ between the two fluids
\begin{equation}
\label{Phase}
{\bm F}^{(r)}_\zeta ({\bm r})=-{\cal G}_{AB} \rho_{\zeta }({\bm r}) \sum_{i, \zeta ' \neq \zeta } w_i \rho_{\zeta '}({\bm r}+{\bm c}_i){\bm c}_i
\end{equation}
is responsible for phase separation~\cite{CHEM09}.  Furthermore, both fluids are also subject to competing interactions whose role is to provide a mechanism for {\it frustration} ($F$) for phase separation~\cite{Seul95}. In particular, we model short range (nearest neighbor, NN) self-attraction, controlled by strength parameters ${\cal G}_{AA,1} <0$, ${\cal G}_{BB,1} <0$), and ``long-range'' (next to nearest neighbor, NNN) self-repulsion, governed by strength parameters ${\cal G}_{AA,2} >0$, ${\cal G}_{BB,2} >0$):
\be\label{NNandNNN}
{\bm F}^{(F)}_\zeta ({\bm r})=-{\cal G}_{\zeta \zeta ,1} \psi_{\zeta }({\bm r}) \sum_{i \in NN} w_i \psi_{\zeta }({\bm r}+{\bm c}_i){\bm c}_i -{\cal G}_{\zeta \zeta ,2} \psi_{\zeta }({\bm r}) \sum_{i \in NNN} w_i \psi_{\zeta }({\bm r}+{\bm c}_i){\bm c}_i ,
\ee
with $\psi_{\zeta }({\bm r})=\psi_{\zeta }[\rho({\bm r})]$ a suitable pseudo-potential function~\cite{SC1,SC2,Sbragaglia07,SbragagliaShan11}. The pseudo-potential is taken in the form originally suggested by Shan \& Chen~\cite{SC1,SC2}:
\be
\label{PSI}
\psi_{\zeta}[\rho_{\zeta}({\bm r})]= \rho_{0} (1-e^{-\rho_{\zeta}({\bm r})/\rho_{0}}).
\ee
The parameter $\rho_{0}$ marks the density value above which non-ideal effects come into play. The prefactor $\rho_{0}$ in (\ref{PSI}) is used to ensure that for small densities the pseudopotential is linear in the density $\rho_{\zeta}$. With the phase separation interactions (\ref{Phase}) we can generate a collection of droplets whose overall stability against coalescence is determined by the stability of the thin films formed between the neighboring droplets. Due to the effect of frustration (\ref{NNandNNN}), a positive disjoining pressure can be achieved~\cite{ourJFM,Sbragaglia12}, which stabilizes the thin films and make the droplets stable against coalescence. As already stressed elsewhere~\cite{ourJFM}, the numerical model possesses two advantages that have been used rarely together. From one side, it gives a realistic structure of the emulsion droplets, like for example the Surface Evolver method~\cite{Surface1,Surface2,Surface3} would do; at the same time, due to its built-in properties, the model gives direct access to equilibrium and out-of-equilibrium stresses~\cite{Sbragaglia12}, including elastic and the viscous contributions. In contrast to other mesoscopic models, such as Durian's bubble model~\cite{Durian97}, our model naturally incorporates the dissipative mechanisms and the interfacial constraints that lead to T1-type plastic events~\cite{SOFT14,ourJFM}.

\section{Numerical Results} \label{sec:results}

\begin{figure}[t]
\begin{center}
\includegraphics[scale=0.4]{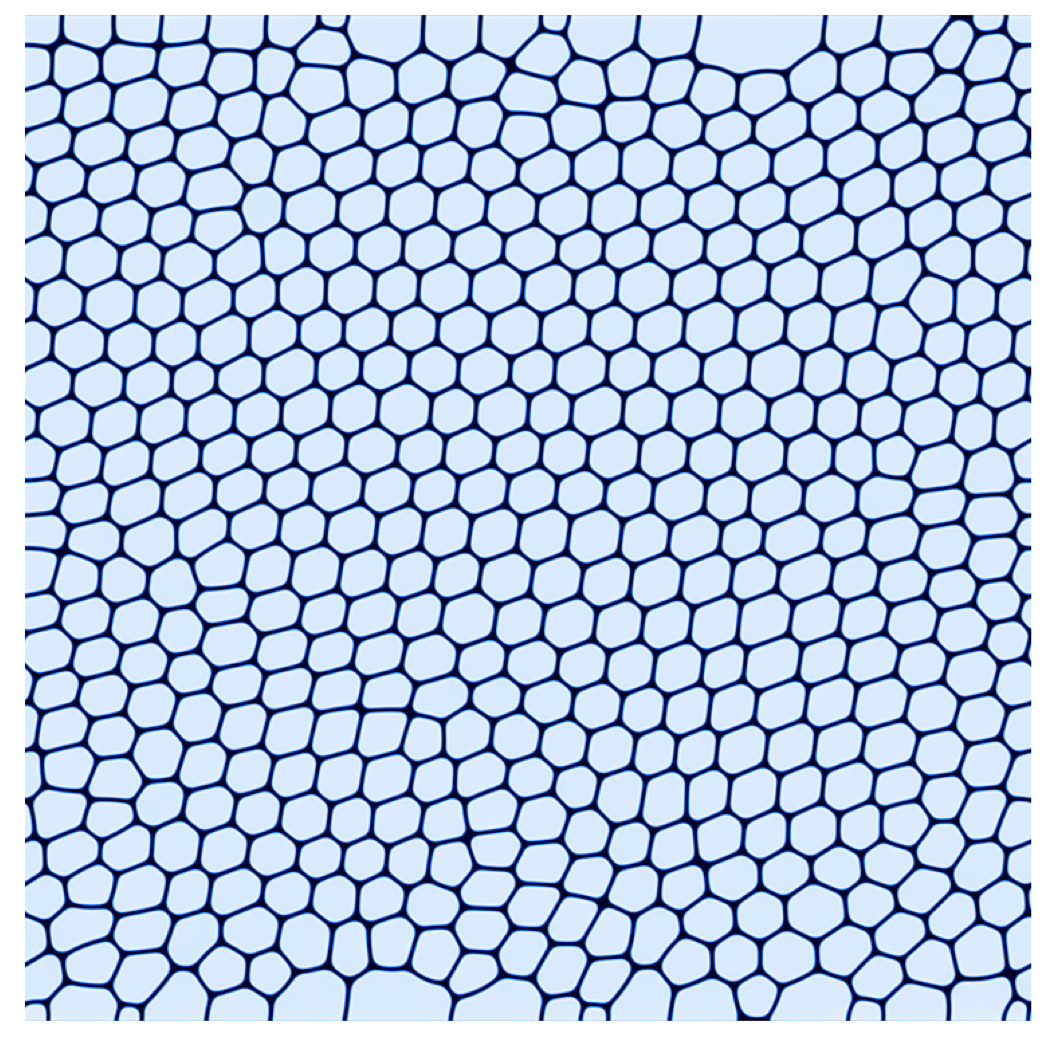}
\includegraphics[scale=0.4]{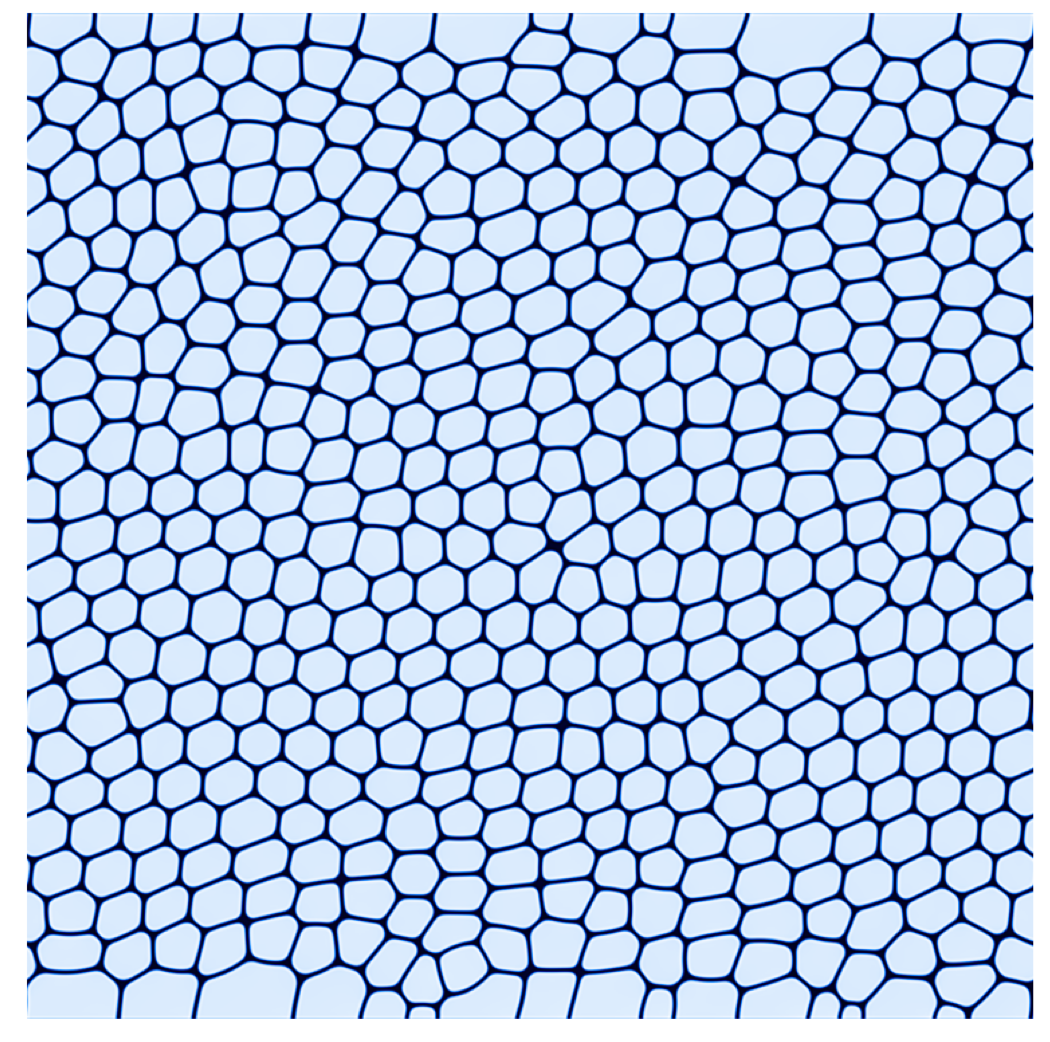}
\caption{We report snapshots of the density field of closely packed droplets in a Couette flow simulated with the lattice Boltzmann models (see Sec. \ref{sec:numerics}). Light (dark) colors refer to regions of space with majority of the dispersed (continuous) phase.  Two sets of simulations have been performed by varying the wall velocity, i.e. imposing a nominal shear rate of: $2 v_w/H=\dot{\gamma}=9.76 \times 10^{-6}$ lbu (left panel) and $\dot{\gamma}=2.92 \times 10^{-5}$ lbu (right panel). Visual inspection reveals, as expected, a larger distortion at larger shears.}
\label{fig:1}
\end{center}
\end{figure}

We studied numerically a planar Couette flow of a 2d dense emulsion confined between two parallel walls in a computational box of $L \times H = 1024 \times 1024$ lattice nodes, with steady velocity at the boundaries $\pm v_w$, and with a volume fraction of the continuous phase $\phi = 7.5 \%$. All the model parameters are exactly the same used in our recent work~\cite{ourJFM}. Two sets of simulations have been performed by varying the wall velocity, which amounts to impose a nominal shear rate of $\dot{\gamma}=2 v_w/H=9.76 \times 10^{-6}$ lbu \footnote{lbu stands for lattice Boltzmann units} (left panel of Fig.~\ref{fig:1}) and $\dot{\gamma}=2.92 \times 10^{-5}$ lbu (right panel of Fig.~\ref{fig:1}). The smaller shear is just above the yielding point. The other shear, instead, is the largest that we can obtain with stable numerical simulations. Stresses are measured as an outcome of the simulations \cite{ourJFM}: we find $\sigma \approx 1.2 \, \sigma_Y$ (for the case $\dot{\gamma}=9.76 \times 10^{-6}$ lbu) and  $\sigma \approx 1.7 \, \sigma_Y$ (for $\dot{\gamma}=2.92 \times 10^{-5}$ lbu).  In both sets of simulations, the parameter $\beta$ has been changed to explore the effect of the wall friction and compare with the theoretical results. To this aim, however, some comments are in order. First, in our derivations we have assumed a HB relation (\ref{eq:fb}) with $a=1$, which is not compatible with the properties of the numerical model, the latter supporting HB rheology \eqref{eq:fb} with $a < 1$ \cite{SOFT14,BenziEPL}. For a quantitative comparison between the numerics and the analytical results of Sec.~\ref{sec:analytics}, we need therefore to determine the ``equivalent'' of the Bingham viscosity $K$ to be used in our theoretical predictions. As a first guess, we can compute such viscosity as $K=\Delta \sigma/\Delta \dot{\gamma}$, where $\Delta \sigma$ ($\Delta \dot{\gamma}$) is just the difference between the two stresses (shear rates) considered in the simulations for $\beta=0$ lbu. We find $K \approx 1.7$ lbu. The other issue concerns the determination of how far/close we are from the yield point, a question that matters in view of the analysis presented in Sec.~\ref{sec:analytics}, in order to decide which analytical prediction to compare with the numerical data. Since the LB simulations refer to a HB fluid \eqref{eq:fb} with $a < 1$ \cite{SOFT14},  the actual stress deviates more slowly from the yield point than it would do in a Bingham fluid under the same shear conditions. Therefore, one heuristically expects a transition from the plastic regime to the fluid regime at relatively smaller values of the stress. This fact will be indeed observed in the numerics.


\begin{figure}[t!]
\begin{center}
\includegraphics[scale=0.4]{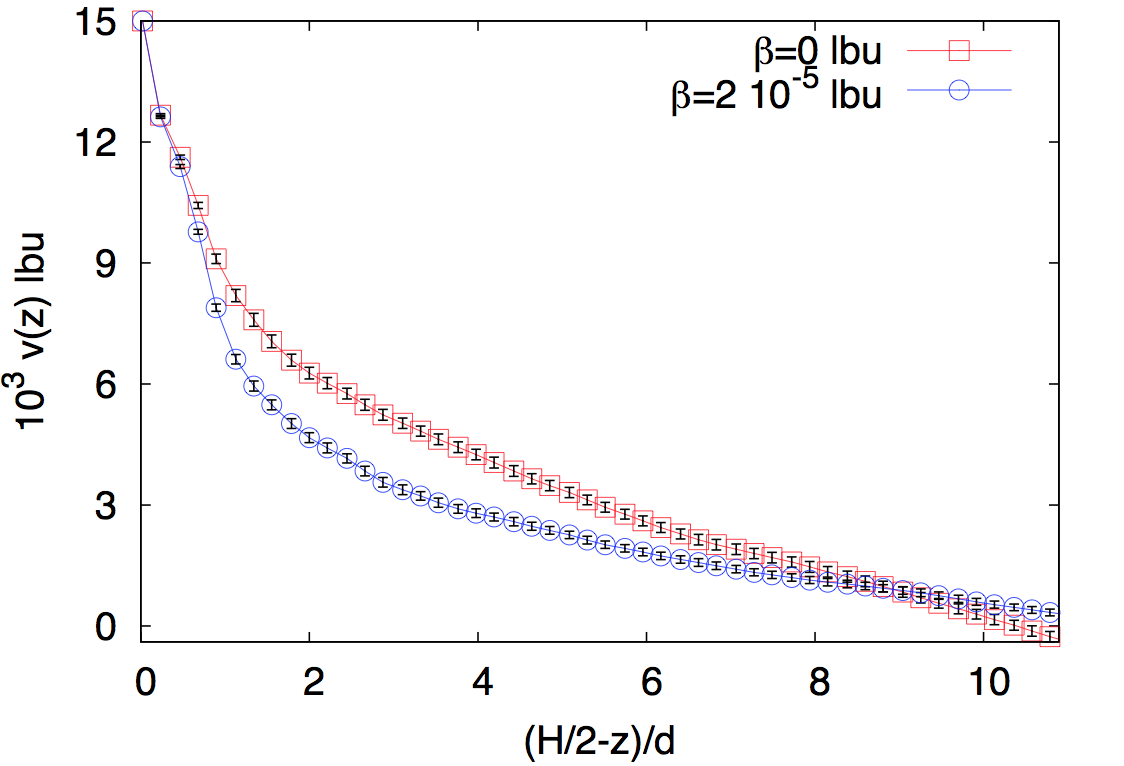}
\includegraphics[scale=0.4]{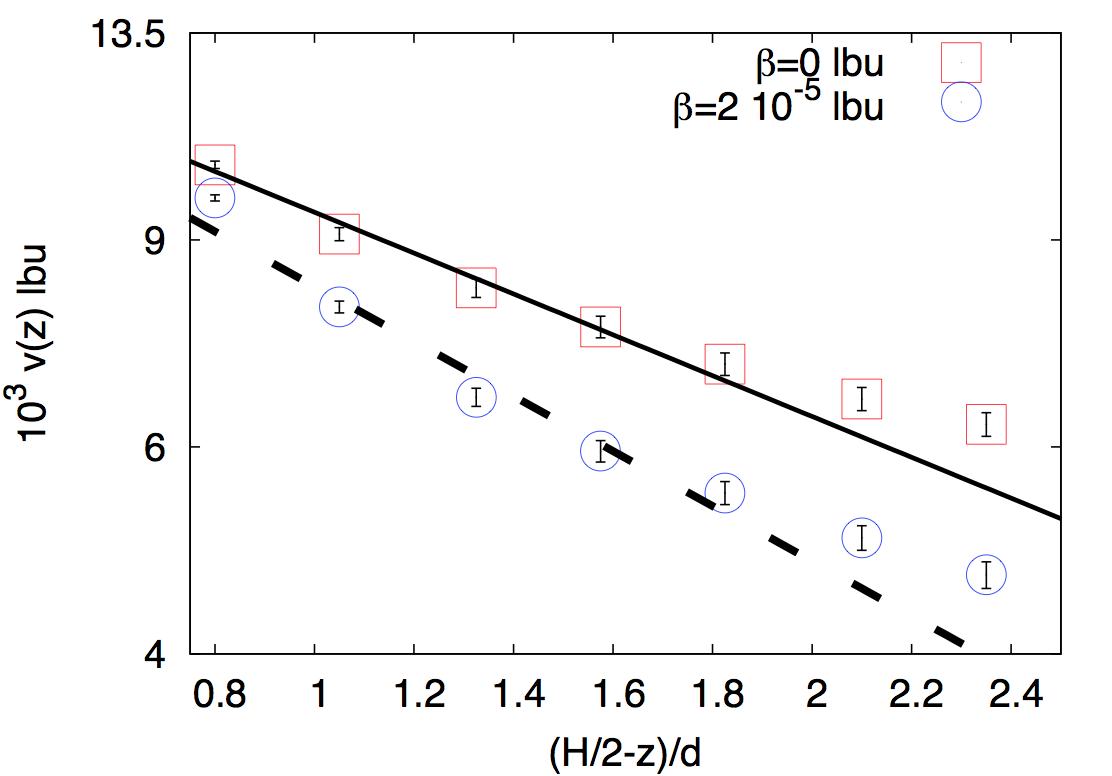}
\caption{Left panel: Velocity profiles for the Couette flow of a 2d soft-glassy material with $\beta=2 \times 10^{-5}$ lbu (corresponding to $L_{\beta} \approx 7.25 \, d$, see text for the details) and $\beta=0$ lbu (without wall friction). The imposed wall velocity is such that the resulting stress is $\sigma \approx 1.7\, \sigma_Y$. Right panel: zoom of a boundary region extended up to a distance $\sim \xi$ from the wall. The solid line is an exponential fit of the frictionless case, which gives a cooperativity length $\xi \approx 2.5 \, d$. With this value, together with $\beta$ and $K$ as input parameters of the simulations, we compute the value of $L_v$ from equation \eqref{eq:loclength}, which imposes the slope of the dashed line. Notice that, consistently with our assumptions, the points start to deviate from the exponential  profiles when the distance from the wall starts to be of the order of the cooperativity length, i.e. $|H/2 - z| \sim \xi$.}
\label{fig:2}
\end{center}
\end{figure}

\begin{figure}[t!]
\begin{center}
\includegraphics[scale=0.5]{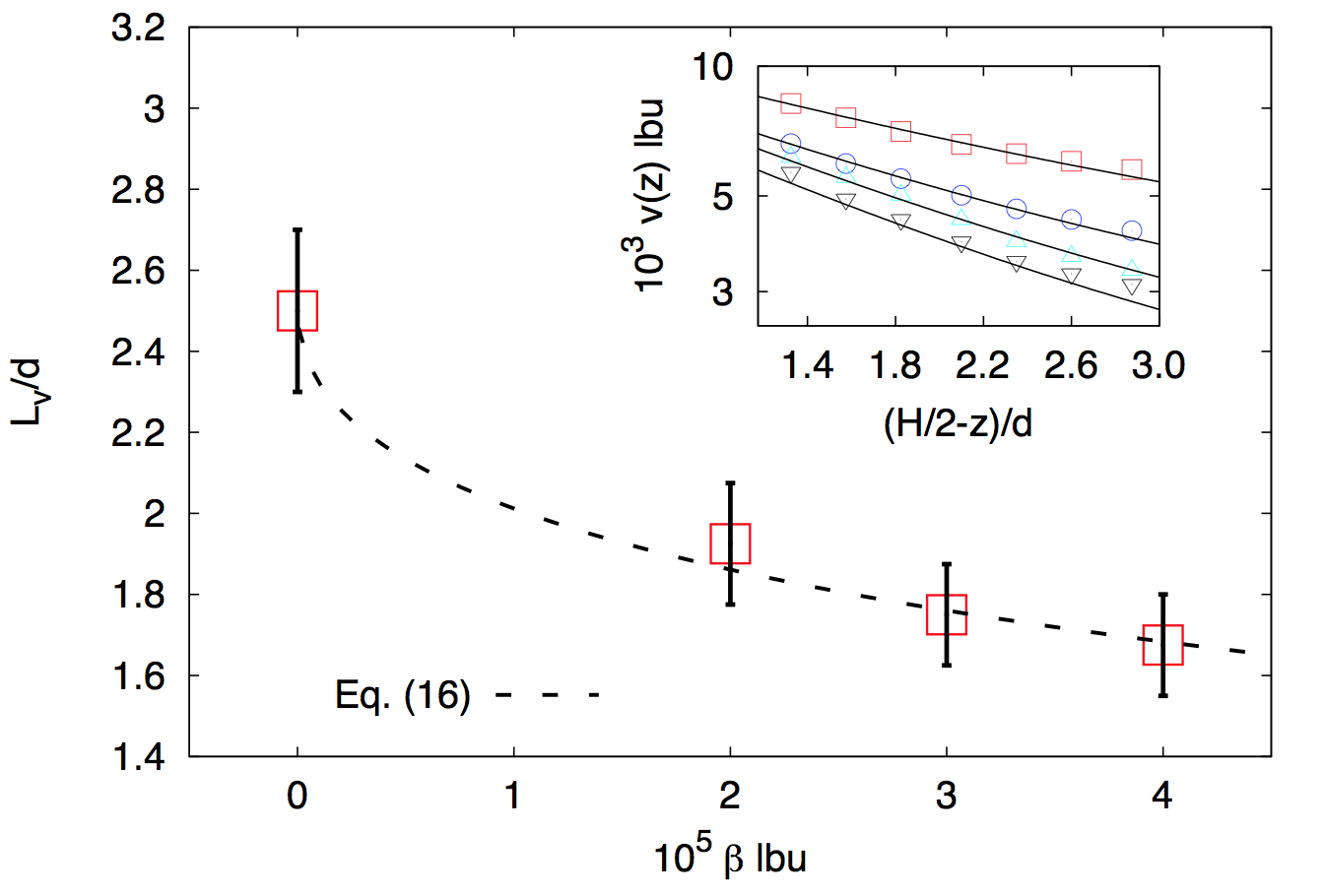}
\caption{Localisation length $L_v$ vs. friction parameter $\beta$. The numerical results (bullets) for $L_v$ are 
compared with the theoretical prediction \eqref{eq:loclength} (dashed line) with the cooperativity length, $\xi \approx 2.5\,d$, obtained via an exponential fit of the velocity profile for $\beta=0$ lbu in the wall proximal region (see also Fig.~\ref{fig:2}). $L_v$ is extracted from a local fit of the velocity profiles (inset) for various $\beta$: $\beta=0$ lbu ($\Box$), $\beta=2 \times 10^{-5}$ lbu ($\circ$), $\beta=3 \times 10^{-5}$ lbu ($\triangle$) and $\beta=4 \times 10^{-5}$ lbu ($\triangledown$). }
\label{fig:3}
\end{center}
\end{figure}

Figure \ref{fig:2} reports the analysis for the case $\sigma \approx 1.7\, \sigma_Y$. In the left panel we plot the velocity profiles without ($\beta = 0$) and with ($L_{\beta} \approx 7.25 \, d$) wall friction. The cooperativity length $\xi \approx 2.5\,d$ is obtained via an exponential fit of the velocity profile for $\beta=0$ lbu in the wall proximal region (solid line in the right panel of Figure \ref{fig:2}). Given the effective plastic viscosity (see above discussion), the friction length $L_{\beta}$ is an input parameter for the simulations. Notice that all the spatial lengthscales are given in units of the mean droplet diameter $d$. In the right panel of figure \ref{fig:2}, we also compare the numerical results in the wall proximal region with the analytical predictions (dashed line) for $\beta \neq 0$ \eqref{eq:velocity2}: the localisation length used, $L_v \approx 1.86 d$, is exactly the value given by equation \eqref{eq:loclength} while the fitted prefactor is $\mathcal{A} \approx 1.25 \times 10^{-4}$ lbu, in reasonable agreement with the analytical prediction $\mathcal{A} = 1.15 \times 10^{-4}$ lbu given by \eqref{eq:Acoeff} (deviating by less than $9 \%$). Fig~\ref{fig:2} actually shows that the ``fluid'' limit of equation \eqref{eq:velocity2} (without any adjustable parameter) is well captured by the LB simulations. To further check the analytical prediction \eqref{eq:loclength}, we performed various numerical simulations at changing the friction parameter. For each $\beta$, the localisation lengths $L_v$'s are extracted from local fits of the velocity profiles (inset of  Fig~\ref{fig:3}). We observe in the main panel of Fig.~\ref{fig:3} that the values of $L_v$ agree very well with the theoretical prediction \eqref{eq:loclength}. We finally turn to the case $\sigma = 1.2\sigma_Y$, close to yield.

\begin{figure}[t!]
\begin{center}
\includegraphics[scale=0.4]{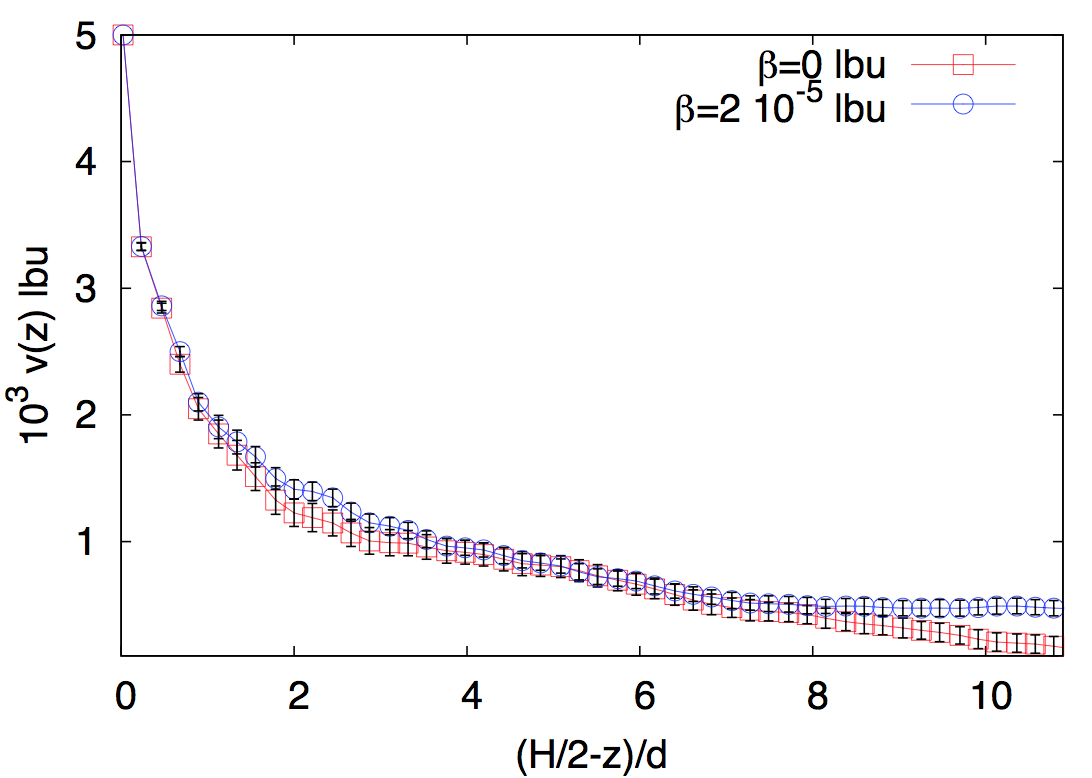}
\includegraphics[scale=0.4]{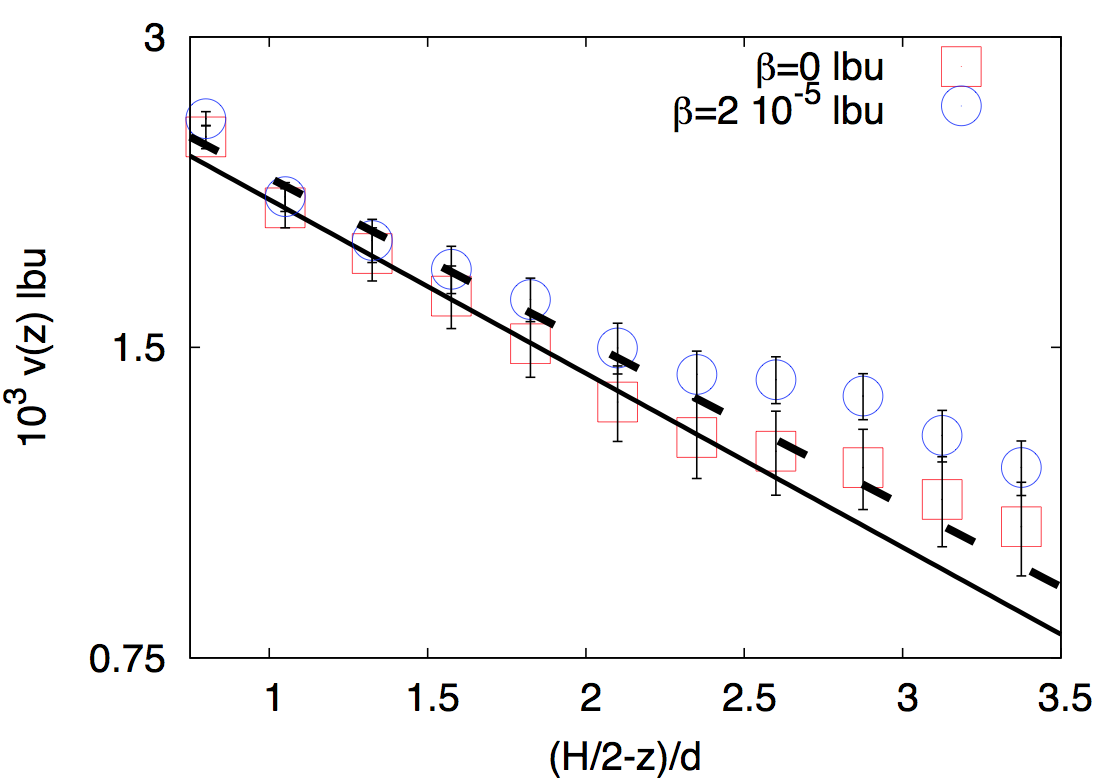}
\caption{Left panel: Velocity profiles for the Couette flow of a 2d soft-glassy meterial with ($\beta=2 \times 10^{-5}$ lbu, corresponding to $L_{\beta} \approx 7.25 \, d$; see text for the details) and without ($\beta=0$ lbu) wall friction. The imposed wall velocity is such that the imposed nominal stress is $\sigma \approx 1.2 \, \sigma_Y$. Right panel: zoom of a boundary region extended up to a distance $\sim \xi$ from the wall. The solid and dashed lines are exponential fits with characteristic lengths $\xi \approx 2.57\,d$ and $L_v \approx 2.74 \, d$ given by equation \eqref{eq:L0}, respectively. Notice that only $\xi$ is fitted, while the localisation length $L_v$ for the case with friction is determined from equation (\ref{eq:L0}) with $K$ as an input parameter for the model (see Sec. \ref{sec:results}).}
\label{fig:4}
\end{center}
\end{figure}

In Fig.~\ref{fig:4} we show, in analogy with Fig.~\ref{fig:2}, the velocity profiles for $\beta=2 \times 10^{-5}$ lbu and the related exponential fits. We could fit the $\beta=0$ case with a cooperativity length $\xi = 2.57\,d$, very close to the one obtained for the stress $\sigma = 1.7\sigma_Y$; with this value (being $L_{\beta} \approx 7.25 \, d$ fixed), equation \eqref{eq:L0} gives a localisation length in reasonable agreement with the numerics.


\section{Conclusions} \label{sec:conclusions}

We have studied both analytically and numerically the Couette flow of a quasi-2d soft-glassy material in a Hele-Shaw geometry. Walls aside the Hele-Shaw cell are sufficiently close to each other to allow visible {\it cooperativity} effects~\cite{Goyon08,Goyon10,Bocquet09}, recently invoked in the literature to rationalize the flow of complex fluids in confined geometries.  Simultaneously, the effect of the confining plates has been modelled by an effective linear friction law, providing {\it shear localisation} for the velocity profiles on lengthscales interfering with the spatial cooperativity~\cite{Goyon08,Goyon10,Geraud13,SOFT14}. For particular rheological flow-curves (Bingham fluids), the problem has been tackled analytically, providing expressions for the two distinct regimes where the material is close to (plastic regime) or well above (fluid regime) the yield point. Other rheo-thinning fluids were also explored with the help of numerical simulations based on lattice Boltzmann models~\cite{SOFT14,ourJFM}, revealing robustness of the analytical findings. Notably, our analysis suggests that the wall friction has different effects in the two regimes: the velocity localisation length is decreased far from yield, while it slightly increases close to yield. Some aspects, however, remain to be further investigated. In the numerics, which simulate a generic HB rheology, the fluid regime is indeed observed to emerge at stresses which are not much larger than the yield stress. This should be attributed to the rheo-thinning character of the fluid, for which the stress grows more slowly with the applied shear than for the Bingham case. It is important to add that this fact, at present, is only supported by numerical simulations. It would be interesting to have complementary experiments with either rheo-thinning and/or Bingham fluids to test the analytical findings at changing the stresses in the material.

Perspectives include more research on the boundary conditions. Even in the absence of slip, which is assumed here and which is realized in practice with rough enough walls, it is not clear what is the boundary condition on the fluidity, and this may affect flow localization. Ultimately, this amounts to reconnect the ``macroscopic" fluidity model to the micromechanics of soft glassy flows \cite{Nicolas13}, and especially how plastic events redistribute the elastic stress in their surroundings, and how this redistribution is affected by the vicinity of the walls \cite{Picard04}.

The authors kindly acknowledge funding from the European Research Council under the EU Seventh Framework Programme (FP7/2007-2013) / ERC Grant Agreement no[279004]. MS and AS gratefully acknowledge M. Bernaschi for computational support.

\end{document}